\documentclass[fp,twocolumn]{jpsj3}
\usepackage{txfonts}

\usepackage{grffile}

\usepackage{graphicx}   % need for figures

\makeatletter
\newcommand\sublabel[2]{%
  \immediate\write\@auxout{\string\newlabel{#1}{{\thefigure #2}{\thepage}}}%
}
\makeatother

\title{Electronic Structure Calculations of CeRh$_{3}$B$_{2}$}

\author{Shotaro Doi$^{1,2}$\thanks{doiexcerpt@gmail.com}, and Munehisa Matsumoto$^2$}
\inst{$^1$Department of Physics, The University of Tokyo, Hongo, Bunkyo, Tokyo 113-0033, Japan \\
$^2$The Institute for Solid State Physics, The University of Tokyo, Kashiwa, Chiba 277-8581, Japan} %\\

\abst{
The electronic structure of the ferromagnetic material CeRh$_{3}$B$_{2}$ is studied by means of first-principles calculations with an emphasis on the treatment of localized $4f$ states around Ce. Via the construction of an effective spin model from electronic structure calculations, we estimated the Curie temperature $T_{\rm C}$ of CeRh$_{3}$B$_{2}$ and found a specific configuration of the electronic structure which explains the exceptionally high measured value of $T_{\rm C} \sim 120$ K as well as the small saturation magnetization and the topology of the Fermi surface. Present advance in the understanding of the subtle nature of the $4f$-electron state in CeRh$_{3}$B$_{2}$ 
that brings about the exceptionally high $T_{\rm C}$ should be also of technological relevance to exploit the utility of Ce in rare-earth permanent magnets.
}

\begin{document}
\maketitle

\section{Introduction}

CeRh$_{3}$B$_{2}$ with the hexagonal CeCo$_{3}$B$_{2}$-type structure has attracted much attention because of its exceptionally high Curie temperature $T_{\rm C} \sim 120$ K \cite{Dhar1981} as a Ce compound without magnetic transition metal elements, despite the very small saturation magnetization $\sim 0.4 \mu_{\rm B} / {\rm f. u.}$ at $4.2$ K\cite{Galatanu2003}. Extensive studies have been reported from both experimental\cite{Ku1980, Dhar1981, Malik1982, Malik1983, Sampathkumaran1985, Yaouanc1998, Alonso1998, Ebihara2000, Galatanu2003, Okubo2003, Givord2004, Kishimoto2004, Yamada2004, Imada2007, Givord2007aug, Givord2007nov, Raymond2010, Ito2014} and theoretical\cite{Takegahara1985,Eriksson1989a,Harima2004,Kono2006,Yamauchi2010} sides for decades, revealing that the ferromagnetism stems from not itinerant Rh-$4d$ bands but localized $4f$-electrons around Ce, that is, the valency of Ce is more likely to be the trivalent configuration of Ce$^{3+}$ with $4f^{1}$@Ce than the tetravalent configuration of Ce$^{4+}$ with $4f^{0}$@Ce. If $T_{\rm C}$'s are estimated for a series of RRh$_{3}$B$_{2}$ (R=Ce, \dots, Gd) with an assumption that the RKKY-type interactions between localized $4f$-electrons around Ce$^{3+}$ realize the ferromagnetic ordering, $T_{\rm C}$ should be maximized for R=Gd from the de Gennes factors of Lanthanide elements. As a matter of fact, the highest $T_{\rm C}$ is observed for R=Ce among the RRh$_{3}$B$_{2}$ series. 
Moreover, while photoemission spectroscopy (PES) and X-ray absorption spectroscopy (XAS) experiments support the electronic configuration of Ce$^{3+}$, the saturation magnetization ($\sim 0.4 \mu_{\rm B} / {\rm f. u.}$) is remarkably reduced from such a large value of the total magnetic moment for free Ce$^{3+}$ ion ($2.14 \mu_{\rm B}$/Ce).
To explain these anomalous trends, many theoretical studies have been performed, e.g., the variational Monte Carlo study\cite{Kono2006}. 

While the electronic structure calculations within the framework of the density functional theory (DFT) have also been reported, studies were restricted to the ground state properties, e.g., the calculation of magnetic moments at $T=0$ K and/or the determination of the Fermi surface (FS)\cite{Takegahara1985,Eriksson1989a,Okubo2003,Harima2004,Yamauchi2010}. Until now, few studies have focused on the estimation of magnetic properties at finite temperature, e.g. $T_{\rm C}$, based on the electronic structure calculations. In the present study, we have estimated such magnetic properties {\it ab-initio} with an emphasis on the treatment of $4f$ states around Ce element.  

The rest of the paper is organized as follows.
In Sec. 2, we describe our methods to realistically describe the electronic structure
involving $4f$ electrons starting from first-principles. Results are given in Sec. 3. Discussions are described along the line of the presentation of the results. Final section is devoted for summary and outlook.
	
\section{Calculation Methods}

The electronic structure calculation was performed for the crystal structure with experimentally measured lattice parameters ($a=5.456$ \AA~ and $c=3.037$ \AA)\cite{Givord2004} using the Korringa-Kohn-Rostoker Green's function method combined with the Coherent Potential Approximation (CPA) on the basis of the DFT. The local (spin) density approximation (L(S)DA) is used to take the exchange-correlation effect into account\cite{Moruzzi1978}. The atomic sphere approximation (ASA) is used for the shape of effective potentials. The effect of the partial wave scattering up to $l_{\rm max}=3$ is considered to treat Ce-$4f$ states as valence states. The effect of relativity is taken into account within the scalar relativistic approximation including the diagonal part of the spin-orbit interaction, $l_{z}s_{z}$.

For the treatment of localized $4f^{1}$-electrons around Ce, care must be taken to overcome the difficulty within the LDA. It is well known that calculations using the LDA give no localized $4f$ bands and form huge unoccupied $4f$ bands whose centers are located at just above the Fermi level, $E_{\rm F}$, which might be regarded as the tetravalent electronic configuration of Ce$^{4+}$. One easy remedy to realize the Ce$^{3+}$ configuration is the so-called ``open core'' approximation, where the localized $4f^{1}$ states are treated as core states whose wave functions are confined within an atomic sphere and the hybridization with other valence bands are switched-off. The other ad hoc approaches based on the LDA scheme such as the LDA+U\cite{Anisimov1997} or the self-interaction correction (SIC)\cite{Perdew1981}, where a certain correction is introduced to realize localized $4f$ states, have also been applied to systems with localized $4f$ states. Without these kinds of localization corrections so that one of $4f$ states around Ce is located at well below the $E_{\rm F}$, it is not possible to realize the energy band structure corresponding to the trivalent electronic configuration of Ce$^{3+}$. 

Studies based on the band structure calculation using the LDA+U scheme were reported for Ce$^{3+}$Rh$_3$B$_2$ by Yamauchi {\it et al.}\cite{Yamauchi2010}. They discussed the electronic configuration of the ferromagnetic ground state and reported quantitative agreement between calculations and experiments for magnetic moments at the ground state with a plausible Ce$^{3+}$ electronic configuration. They also reported that the experimentally identified topology of the FS of CeRh$_3$B$_2$, which is similar to the one of the LaRh$_3$B$_2$, is reproduced from the band structure calculations combined with a certain artificial correction that the Ce-$d$ and $f$ levels are shifted upward by a few tenth of Ry from the ones obtained using the LDA.

Concerning the above findings on the FS, we explored a condition to realize the calculated value of the Curie temperature $T_{\rm C}^{\rm calc.}$ which is comparable with the experimental value, $T_{\rm C}^{\rm expt.} \sim 120$ K, with/without artificial shifts for Ce-$d$ and $f$ levels, shifted upwards by 0.13 Ry and 0.12 Ry respectively\cite{Yamauchi2010}, using the localized correction schemes for localized $4f$ states via the evaluation of the effective exchange interaction, $J_{ij}$'s, explained as follows. 

\subsection{Localized correction scheme}
To obtain the electronic structure of Ce$^{3+}$Rh$_{3}$B$_{2}$, we used several correction schemes for $4f$ states $\left( l=3; m=-\left| l \right|, \cdots, \left| l \right| \right)$: the ``open core'' approximation, the LDA+U and the pseudo SIC method, based on the notion of the orbital dependent potentials, $v_{{\rm eff}}^{l=3; m}$, which is generally expressed as
\begin{align}
	v_{{\rm eff}}^{l=3; m}(\mathbf{r}) = v_{\rm eff}^{\rm LDA}(\mathbf{r}) +\Delta v_{l=3; m}(\mathbf{r}),
\end{align}
where $v_{\rm eff}^{\rm LDA}$ is the effective potential within the LDA and $\Delta v_{l=3; m}$ is the correction term for a selected orbital $(l=3; m)$.

\subsubsection{``Open core'' approximation}
In the ``open core'' approximation, 
Ce-$4f$ states are treated as core states by setting $l_{\rm max}=2$, resulting in the removal of $4f$ bands from valence states.
Furthermore, the effect of the SIC for the localized core $4f$ states are taken into account as
\begin{align}
	\Delta v_{l=3; m}(\mathbf{r}) =  - V_{\rm HXC}\left( \left| \phi_{l=3; m}^{\rm core}(\mathbf{r}) \right|^{2} ; 1 \right),
\end{align}
where $\phi_{l=3; m}^{\rm core}$ is the core $4f$ wave function normalized within the atomic sphere, and $V_{\rm HXC}\left( \rho\left( \mathbf{r} \right) ; 1 \right)$ is the Hartree plus exchange-correlation potential for the charge density $\rho\left( \mathbf{r} \right)$ with the full spin-polarization. Within the spherical potential approximation such as the ASA, the $m$ dependence is dropped and it is sufficient to solve a core electron problem where one electron is confined within an $4f$ core orbital at the Ce site. The ``open core'' treatment is thus supposed to be one of the easiest prescription to form the Ce$^{3+}$ electronic structure. While the effect of the hybridization between the core Ce-$4f$ states and other valence states are completely discarded in the ``open core'' approximation, in the following correction schemes, the localized Ce-$4f$ states are treated as valence states, where the hybridization between the localized Ce-$4f$ states and other valence states are still maintained.

\subsubsection{LDA+U-AL}
 The LDA+U total energy functional is generally expressed as
 \begin{align}
 	E &= E_{\rm LDA} + E^{U},
\end{align}
where $E_{\rm LDA}$ is the usual LDA total energy functional and $E^{U}$ is the correction to the LDA energy.
Accordingly, the correction term of the orbital dependent potential for $(l=3; m)$ in the LDA+U scheme is given as
\begin{align}
	\Delta v_{l=3; m}(\mathbf{r}) =  - v_{l=3; m}^{U}.
\end{align}
There are wide variety of LDA+U functionals depending on a choice of $E^{U}$. The commonly used form is the so-called ``atomic limit''(AL) type (named as the LDA+U-AL)\cite{Czyzyk1994,Dudarev1998,Shick1999}.
In the LDA+U-AL scheme, the correction term of the orbital dependent potential is given as
\begin{align}
	v_{l=3; m}^{U}=(U-J)\left( n_{l=3; m} - \frac{1}{2} \right),
\end{align}
where $U$ and $J$ are the adjustable parameters in the LDA+U scheme.

\subsubsection{LDA+U-SIC}
Recently, another LDA+U expression that was designed to reproduce the correct total energy of an isolated hydrogen atom regarding its SIC appropriately, was proposed by Seo (named as the LDA+U-SIC)\cite{Seo2007}.
In the LDA+U-SIC scheme, the correction term of the orbital dependent potential is given as
\begin{align}
	v_{l=3; m}^{U}=(U-J)n_{l=3; m}.  \label{eq: LDA+U-SIC}
\end{align}

\subsubsection{Pseudo SIC}
A computationally efficient way to consider the SIC for selected orbitals, so-called the pseudo self-interaction-correction (pSIC), was proposed by Filippetti {\it et al.}\cite{Filippetti2003, Pemmaraju2007, Filippetti2009, Filippetti2011} .
In the pSIC scheme, the orbital dependent part of the effective potential is given as
\begin{align}
	\Delta v_{l=3; m}(\mathbf{r}) = -  \alpha n_{l=3; m} V_{\rm HXC}\left( \left| \phi_{l=3; m}(\mathbf{r}) \right|^{2} ; 1 \right),  \label{eq: pSIC}
\end{align}
where $\alpha$ is an empirical scaling factor which was set to $1/2$ in the original article, $n_{l=3; m}$ is the number of occupied $4f$ electron for a $\left( l=3; m\right)$ state, $\phi_{l=3; m}$ is the corresponding wave function normalized within the atomic sphere.

We applied the above correction schemes, the ``open core'' approximation, as well as the three correction schemes, the LDA+U-AL, the LDA+U-SIC and the pSIC for the localized $4f$ states around Ce$^{3+}$. Within all the latter three correction schemes, the localized 4f states below the $E_{\rm F}$ are shifted downward. And the contrasting feature is evident: whereas the LDA+U-AL correction shifts upward the $4f$ states for $n_{l=3; m} < 1/2$, the LDA+U-SIC and the pSIC corrections do nothing for unoccupied $4f$ states above the $E_{\rm F}$.

\subsection{Electronic configuration of localized $4f$ electrons for Ce$^{3+}$}
\label{electronic configuration}
Kono {\it et al.} discussed the electronic configuration of localized $4f$ electrons for Ce$^{3+}$ at the ferromagnetic ground state\cite{Kono2006}. From the crystal-field theory, in the hexagonal symmetry, the ground-state multiplet with $J=5/2$ for the lowest $4f^{1}$ state($L=3$ and $S=1/2$) is split into three Kramers doublets $| \pm J_{z} \rangle$ each corresponding to $| J_{z} \rangle =5/2, 3/2$, and $1/2$ with the descending order with respect to the energy level. Concerning the in-plane magnetocrystalline anisotropy and the fact that the high excitation energy of 220 K between $| \pm 3/2\rangle$ and $| \pm 1/2\rangle$ is far above the $T_{\rm C}$\cite{Yamada2004}, the ground state wave functions along the $x$-axis, $\Phi_x$, was proposed as
\begin{align}
	\Phi_x=| 1/2\rangle + | -1/2\rangle,
\end{align}
where 
\begin{align}
	|\pm 1/2\rangle = \pm \sqrt{\frac{4}{7}} | \pm 1, \mp 1/2\rangle \mp  \sqrt{\frac{3}{7}} | 0, \pm 1/2\rangle,
\end{align}
and $| L_{z}, S_{z}\rangle$ stands for a basis of $L=3$ and $S=1/2$.

In the present study, we choose $\Phi_x$ as the electronic configuration of localized $4f$ states for Ce$^{3+}$ to perform the electronic structure calculations for the ferromagnetic ground state. However, it is difficult to consider the electronic configuration comprising of the superposition of multiple states as of $\Phi_x$. To overcome the difficulty, we mimicked the electronic configuration using the pseudo alloying approach: we consider a system with a Ce sublattice at which two types of Ce atoms corresponding to two different states randomly occupy, each with a certain concentration following the coefficients in the expression of $\Phi_x$, $4/7$ for $| -1,1/2\rangle$ and $3/7$ for $| 0,1/2\rangle$ per the majority spin-channel. The electronic structure calculations for the pseudo alloying state is performed using the CPA.

\subsection{Effective exchange interaction $J_{ij}$}
For a ferromagnetic ground state, the effective exchange interaction between magnetic atoms, $J_{ij}$, are evaluated using the prescription proposed by Liechtenstein {\it et al.}\cite{Liechtenstein1987}. Once the effective classical Heisenberg-type model is constructed by evaluating $J_{ij}$'s for a magnetic state, which could be conceived as a model that describes the effect of the transverse spin fluctuations for a magnetic itinerant electron system, it is possible to estimate the magnetic transition temperature using whatever acceptably good method to solve
the classical Heisenberg model, e.g., the mean-field approximation (MFA). We estimated the $T_{\rm C}$ of CeRh$_{3}$B$_{2}$ within the MFA via the evaluation of $J_{ij}$'s for several treatments of 4f states around Ce: the LDA for Ce$^{4+}$; the ``open core'' approximation, the LDA+U-AL, the LDA+U-SIC and the pSIC for Ce$^{3+}$.

\section{Results and Discussion}
\subsection{Ce$^{4+}$ within the LDA}
Fig. \ref{fig: band Ce$^{4+}$} shows the calculated band dispersion of the majority spin channel for Ce$^{4+}$Rh$_{3}$B$_{2}$ within the LDA. No $4f$ states are localized and these states lie at just above the $E_{\rm F}$. We confirmed that parts of Ce-$d$ and $f$ bands cross the $E_{\rm F}$ without artificial shifts for these levels, which contradict the topology of the FS deduced from the de Haas$–$van Alphen (dHvA) experiments\cite{Okubo2003, Yamauchi2010}. 

The calculated spin and orbital moments on the Ce site are 0.22 $\mu_{\rm B}$ and  -0.07 $\mu_{\rm B}$.  The calculated $J_{ij}$ between the nearest Ce atoms along the $c$-axis, $J_{\rm Ce, Ce}\sim 0.7$ meV was obtained and the rest of $J_{ij}$'s between all the other pairs were found to be negligible. Correspondingly, the calculated value of $T_{\rm C}^{\rm calc.}\sim28$ K was obtained by the MFA, which is much lower than $T_{\rm C}^{\rm expt.}$.

\begin{figure}[hbtp]
 \centering
 \includegraphics[clip,width=1.0\linewidth]{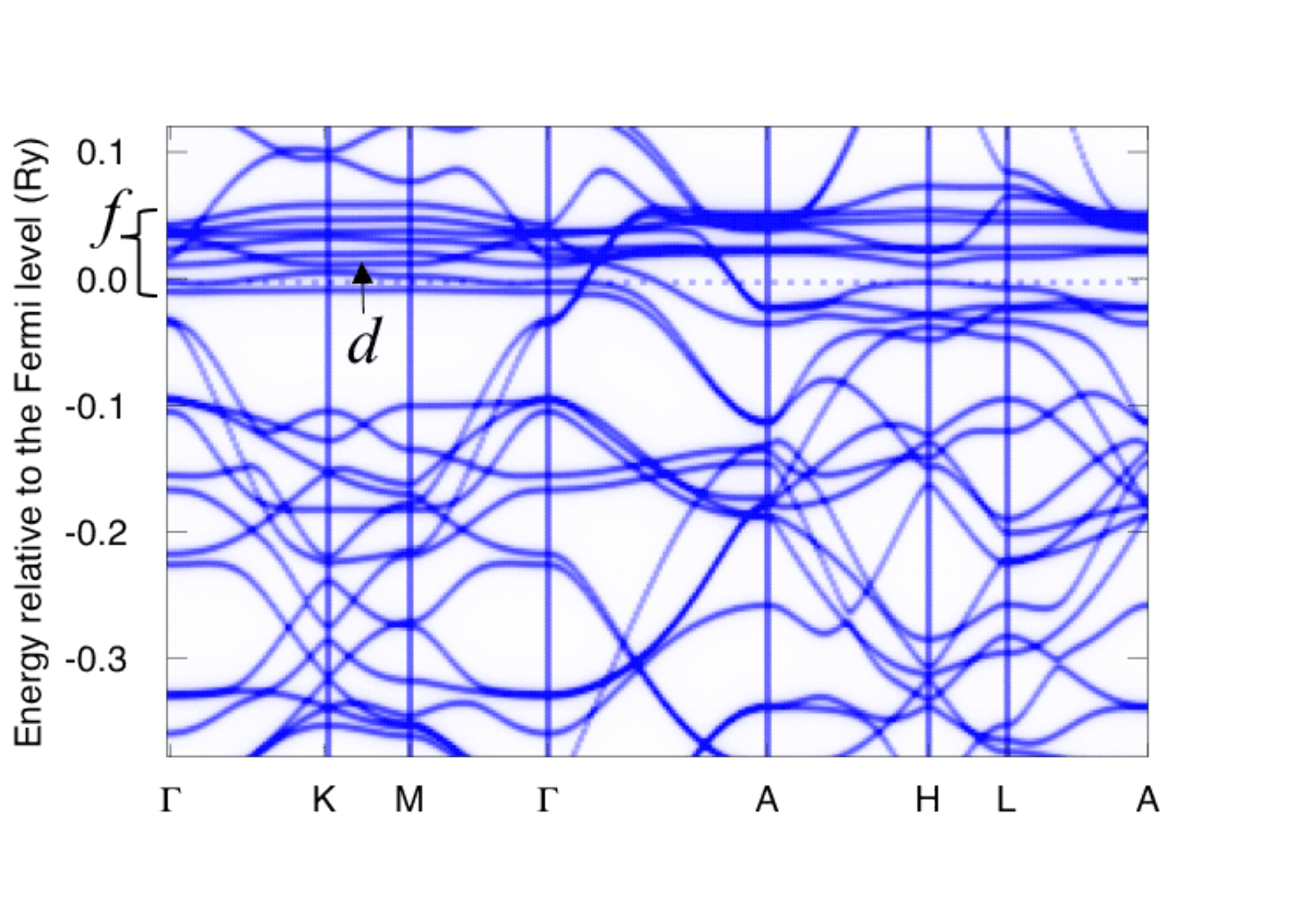}
 \caption{Electronic band structure of Ce$^{4+}$Rh$_{3}$B$_{2}$ within the LDA scheme for the majority spin channel}
 \label{fig: band Ce$^{4+}$}
\end{figure}

\subsection{Ce$^{3+}$ with the localization correction schemes}
\subsubsection{``Open core'' approximation}
Fig. \ref{fig: band Ce$^{3+}$ open core} shows the calculated band dispersion of the majority spin channel for Ce$^{3+}$Rh$_{3}$B$_{2}$ in the ``open core'' approximation. No $4f$ bands appear in the figure because the localized $4f$ states around Ce are treated as core states. The band structure is very similar to the one of LaRh$_{3}$B$_{2}$ and artificial shifts for Ce-$d$ levels are still inevitable to reproduce the measured FS to avoid appearance of bands at the FS found in regions $\Gamma$-K and M-$\Gamma$, which is attributed to the inappropriate position of Ce-$d$ bands\cite{Okubo2003}. 

The calculated spin and orbital moments on the Ce site are 1.06 $\mu_{\rm B}$ and -0.03 $\mu_{\rm B}$, including 1 $\mu_{\rm B}$ from the full polarization of the 4f core state.  The calculated value of $J_{\rm Ce, Ce}\sim 0.5$ meV was obtained, and as is the case for Ce$^{4+}$ within the LDA, the rest of $J_{ij}$'s were negligibly small. The calculated value of $T_{\rm C}^{\rm calc.}\sim 18$ K obtained by the MFA is also much lower than $T_{\rm C}^{\rm expt.}$. 

It is straightforward to evaluate $J_{ij}$'s for R=Gd within the ``open core'' approximation. Results for R=Ce$^{3+}$ and Gd are shown in Table \ref{table: Tc open core}. We used the experimental lattice parameters for R=Gd\cite{Ku1980}. $T_{\rm C}^{\rm calc.}$ for R=Ce$^{3+}$ is lower than that for R=Gd as is the case with the description based on the RKKY-type interaction model, resulting in contradiction to experiments. It is plausible that the RKKY-type model and the ``open core'' approximation is similar in terms of the indirect contribution between localized $4f$ electrons around Ce for $J_{\rm Ce, Ce}$ mediated via the sea of conduction electrons. There are differences from the simple RKKY-type model: $T_{\rm C}^{\rm calc.} \sim 18$ K for R=Ce$^{3+}$ is higher than that of the RKKY-type model by two orders of magnitude, and $T_{\rm C}^{\rm calc.}  \sim 44$ K for R=Gd is still much lower than the experimental value, which would stem from the complex electronic band structure than the one assumed in the simple RKKY-type model.

\begin{figure}[hbtp]
 \centering
 \includegraphics[clip,width=1.0\linewidth]{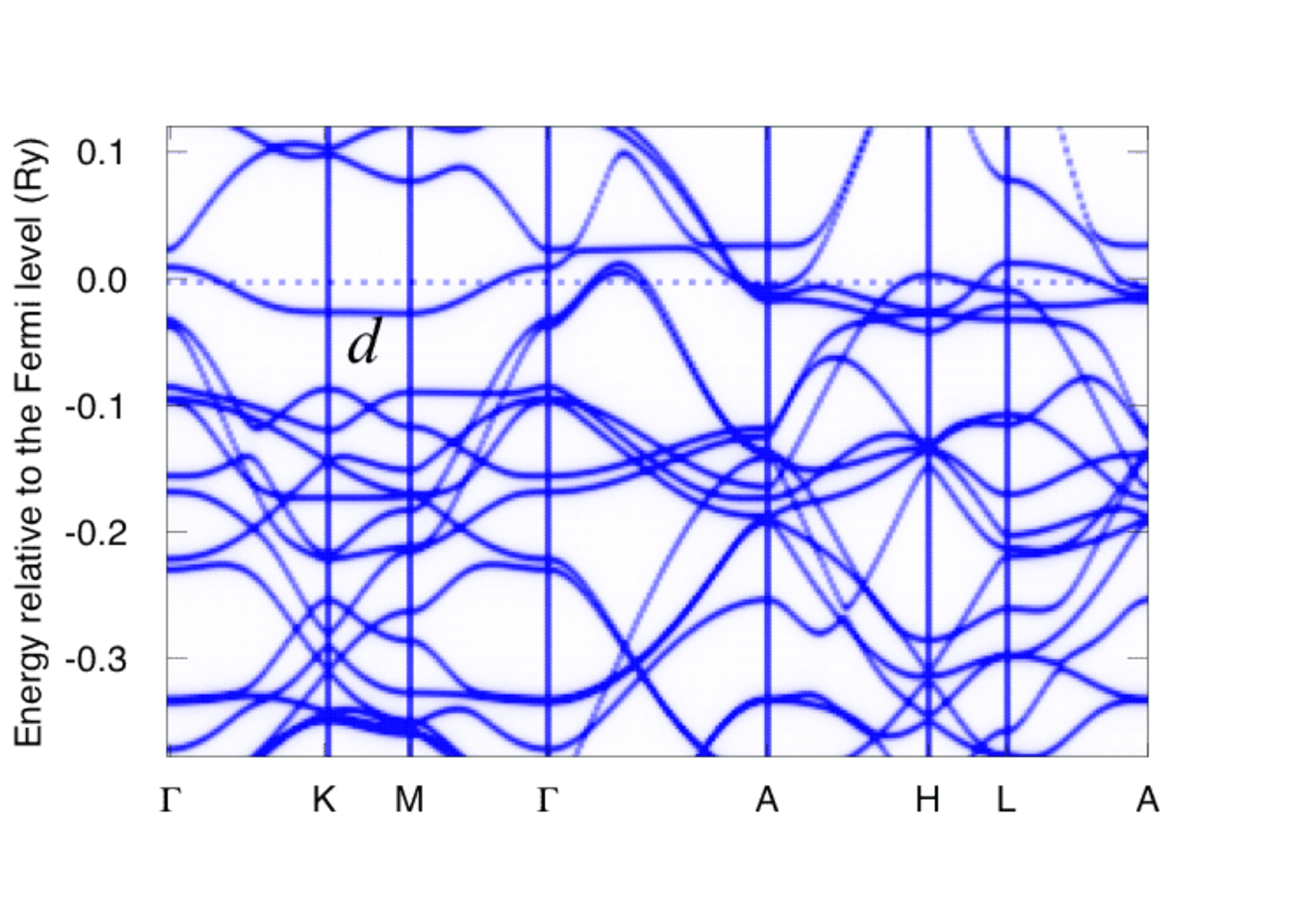}
 \caption{Electronic band structure of Ce$^{3+}$Rh$_{3}$B$_{2}$ in the ``open core'' approximation for the majority spin channel}
 \label{fig: band Ce$^{3+}$ open core}
\end{figure}

\begin{table}[hbtp]
	\begin{center}
		\caption{Calculated $T_{\rm C}$ for R=Ce$^{3+}$ and Gd using the ``open core'' approximation}
		\begin{tabular}{ ccc} \hline
			R= & Ce$^{3+}$ & Gd \\ \hline \hline
			$T_{\rm C}^{\rm calc.}$ [K] & $\sim 18$   & $\sim 44$ \\
			$T_{\rm C}^{\rm expt.}$ [K] & $\sim 120$ & $\sim 90$ \\ \hline
		\end{tabular}
		\label{table: Tc open core}
	\end{center}
\end{table}

\subsubsection{LDA+U-AL}
\label{LDA+U-AL}
\begin{figure}[hbtp]
 \centering
 \includegraphics[clip,width=1.0\linewidth]{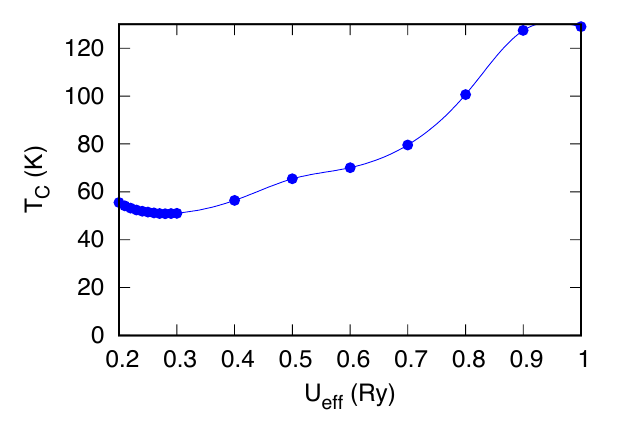}
 \caption{Calculated $T_{\rm C}$ as a function of $U_{\rm eff}$ using the LDA+U-AL scheme}
 \label{fig: Tc LDA+U-AL}
\end{figure}

Fig. \ref{fig: Tc LDA+U-AL} shows the calculated $T_{\rm C}$ as a function of $U_{\rm eff}=U-J$ parameter using the LDA+U-AL scheme. For $U_{\rm eff} < 0.2$ Ry, the self-consistent field (SCF) solution that satisfies the electronic configuration with the condition imposed in Sec. \ref{electronic configuration} could not be obtained. While $T_{\rm C}^{\rm calc.}\sim 120$ K were obtained for $U_{\rm eff}\sim 1$ Ry, the parameter region, which particularly results in deep $4f$ states, is supposed to be unrealistic concerning the fact that the typical value of $U \sim 0.4$ Ry for Ce$^{3+}$ compounds has been deduced from calculations referring to the PES experiments\cite{Cococcioni2005}. 

\begin{figure}[hbtp]
	\centering
        	\includegraphics[clip,width=1.0\linewidth]{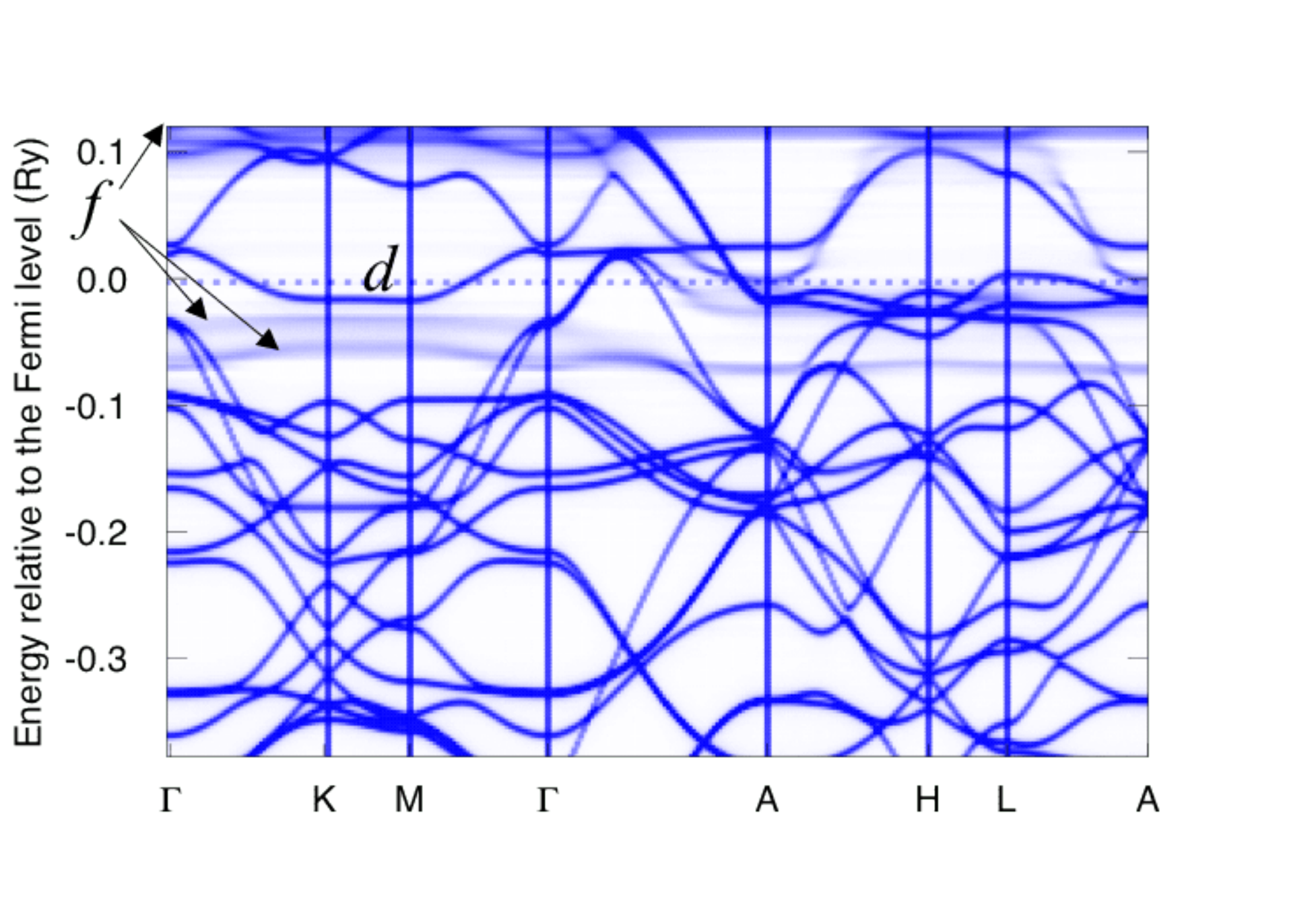}	
	\caption{Electronic band structure of Ce$^{3+}$Rh$_{3}$B$_{2}$ for the majority spin channel using the LDA+U-AL scheme for $U_{\rm eff}=0.2$ Ry}
	\label{fig: band LDA+U-AL}
\end{figure}

Fig. \ref{fig: band LDA+U-AL} shows the calculated band dispersion of the majority spin channel for Ce$^{3+}$Rh$_{3}$B$_{2}$ with $U_{\rm eff}=0.2$ Ry, where we obtained the SCF solution with the aimed electronic configuration for a $U_{\rm eff}$ parameter as small as possible. We obtained the band dispersion for the pseudo alloy systems by plotting the Bloch spectral functions for the disordered alloys within the CPA\cite{Faulkner1980}. The $4f$ bands with narrow dispersion located at just above the $E_{\rm F}$ within the LDA are split into two groups after the LDA+U correction: unoccupied bands at $\sim E_{\rm F}+0.1$ Ry and occupied bands at $\sim E_{\rm F}-0.05$ Ry. Moreover, the smearing of $4f$ bands are observed because the pseudo alloying approach for the mixture of two Ce states leads to the smearing of bands attributing to Ce.

\subsubsection{LDA+U-SIC}
\begin{figure}[hbtp]
 \centering
 \includegraphics[clip,width=1.0\linewidth]{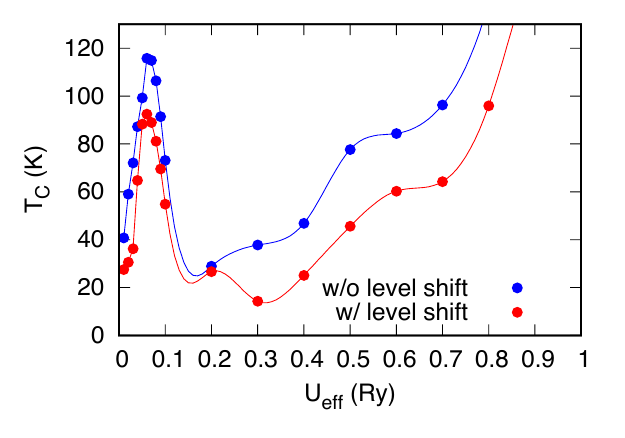}
 \caption{Calculated $T_{\rm C}$ as a function of $U_{\rm eff}$ using the LDA+U-SIC scheme}
 \label{fig: Tc LDA+U-SIC}
\end{figure}

Fig. \ref{fig: Tc LDA+U-SIC} shows the calculated $T_{\rm C}$ as a function of $U_{\rm eff}$ parameter using the LDA+U-SIC scheme with/without artificial level shifts for Ce-$d$ and $f$ states. While $T_{\rm C}$ is a increasing function of $U_{\rm eff}$ in a region with large $U_{\rm eff}$ ($\to 1$ Ry), we can safely omit the results for the reason explained in Sec.\ref{LDA+U-AL}. Without the level shifts, the maximum of the calculated value of $T_{\rm C}^{\rm calc.}\sim 114$ K is obtained for $U_{\rm eff}=0.06$ Ry in a region with relevant $U_{\rm eff}$ ($\leq$ $\sim 0.4$ Ry: so as not to bring too deep occupied $4f$ states), which is comparable to $T_{\rm C}^{\rm expt.}$. We also found a systematic tendency that the calculated value of $T_{\rm C}^{\rm calc.}$'s are decreased in the whole range of $U_{\rm eff}$ with the level shifts. The calculation with the level shifts gave the maximum of the calculated value of $T_{\rm C}^{\rm calc.}\sim 92.5$ K for $U_{\rm eff}=0.06$ Ry as is the case without the level shifts. 

The calculated spin and orbital moments on the Ce site without (with) the level shifts for $U_{\rm eff}=0.06$ Ry are,  0.72 (0.70) $\mu_{\rm B}$ and -0.45 (-0.51) $\mu_{\rm B}$ for $|-1,1/2\rangle$, 0.81 (0.71) $\mu_{\rm B}$ and -0.09 (-0.07)$\mu_{\rm B}$ for $|0,1/2\rangle$, respectively, leading to the averaged value of spin and orbital moments as 0.77 (0.71) $\mu_{\rm B}$ and -0.24 (-0.26) $\mu_{\rm B}$. The calculated value of the total magnetic moment of 0.53 (0.45) $\mu_{\rm B}$ is comparable to the measured value of $\sim 0.4 \mu_{\rm B}$.

\begin{figure}[hbtp]
	\begin{minipage}{1.0\linewidth}
		\centering
        		\includegraphics[clip,width=1.0\linewidth]{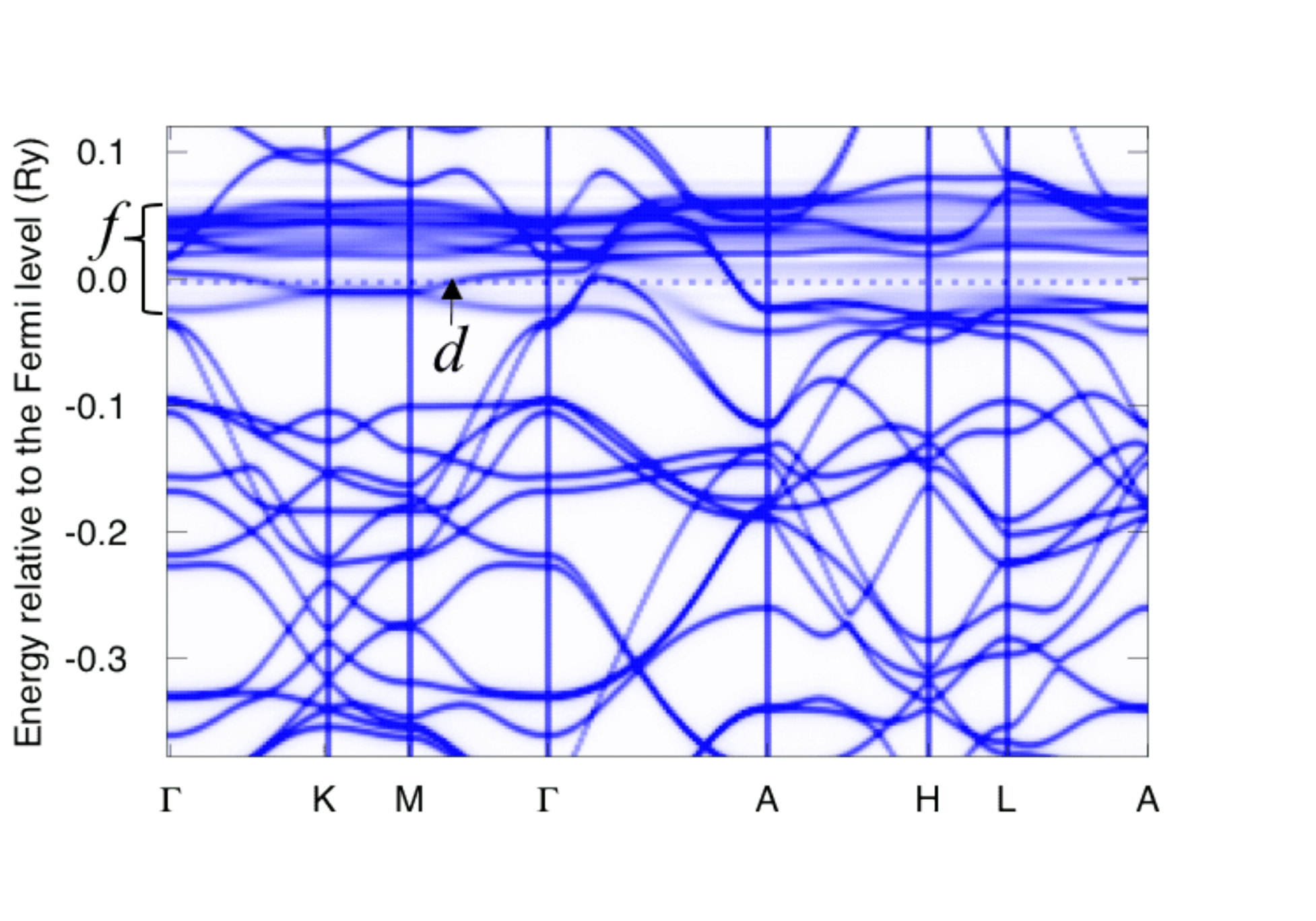}
		(a) without level shifts
	\end{minipage}\\
	\begin{minipage}{1.0\linewidth}
		\centering
		\includegraphics[clip,width=1.0\linewidth]{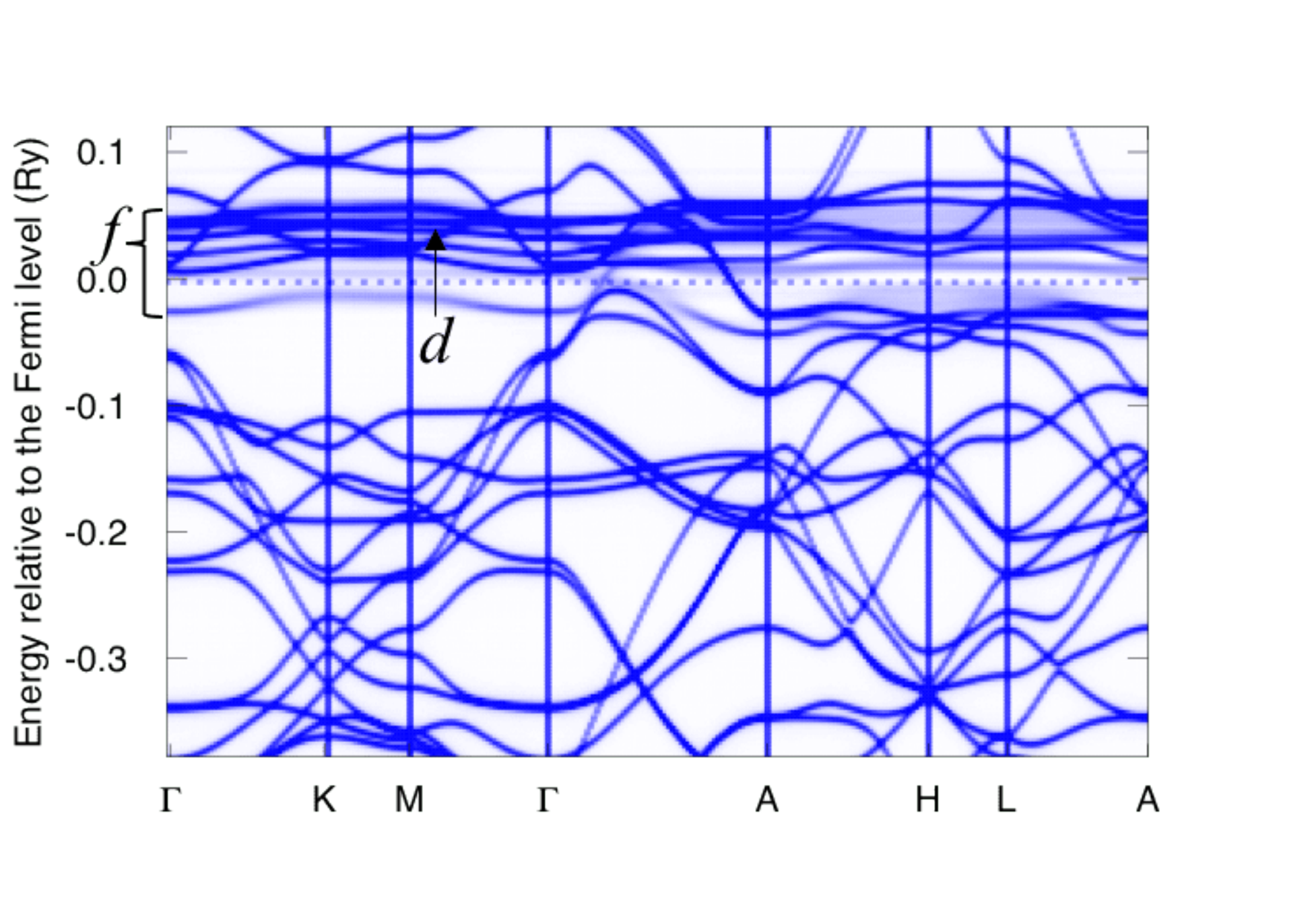}
		(b) with level shifts
	\end{minipage}
	\caption{Electronic band structure of Ce$^{3+}$Rh$_{3}$B$_{2}$ for the majority spin channel using the LDA+U-SIC scheme for $U_{\rm eff}=0.06$ Ry at which the calculated  $T_{\rm C}$ is maximized}
	\label{fig: band LDA+U-SIC}
	\sublabel{fig: band LDA+U-SIC without shifts}{(a)}
	\sublabel{fig: band LDA+U-SIC with shifts}{(b)}
\end{figure}

Fig. \ref{fig: band LDA+U-SIC} shows the calculated band dispersion of the majority spin channel for Ce$^{3+}$Rh$_{3}$B$_{2}$ with $U_{\rm eff}=0.06$ Ry, where the calculated $T_{\rm C}$ is maximized, with/without artificial level shifts. Again, owing to the pseudo alloying treatment for the mixture of the two Ce$^{3+}$ states, we observed the smearing of band dispersion, especially around the localized $4f$ bands with narrow dispersion whose centers locate at just below the $E_{\rm F}$. Without the level shifts, the crossing of Ce-$d$ bands and the $E_{\rm F}$ are found in regions, $\Gamma$-K and M-$\Gamma$, which apparently contradicts the measured topology of the FS\cite{Yamauchi2010}. The undesirable situation is relieved with the level shifts as shown in Fig. \ref{fig: band LDA+U-SIC with shifts}.

Within the LDA+U-SIC scheme, we found a typical band structure which enhances $J_{\rm Ce, Ce}$ and the corresponding calculated value of $T_{\rm C}$ that is comparable to the measurements, where the localized $4f$ states locate at just below the $E_{\rm F}$, characteristically.
Recent resonant PES experiment reported that the excitation $4f$ spectrum has a peak at $\sim 0.12$ eV below the $E_{\rm F}$ with a width of $\sim 0.4$ eV, which can be attributed to the localized $4f$ state around Ce\cite{Imada2007}. The value of $U_{\rm eff}=0.06$ Ry and the corresponding position of the localized $4f$ bands are supposedly in good agreement with the resonant PES measurement.

%\begin{figure}[hbtp]
%	\centering
% 	\subfigure[without level shifts]{
%       		\centering
%        		\includegraphics[clip,width=1.0\linewidth]{pdf/cerh3b2_ls_pu_3_4_0.06_up_spc.pdf}
%		\label{fig: band LDA+U-SIC without shifts}
%	}
% 	\subfigure[with level shifts]{
%		\centering
%		\includegraphics[clip,width=1.0\linewidth]{pdf/shft_cerh3b2_ls_pu_3_4_0.06_up_spc.pdf}
%		\label{fig: band LDA+U-SIC with shifts}
%	}
%	\caption{Electronic band structure of Ce$^{3+}$Rh$_{3}$B$_{2}$ for the majority spin channel using the LDA+U-SIC scheme for $U_{\rm eff}=0.06$ Ry at which the calculated  $T_{\rm C}$ is maximized}
%	\label{fig: band LDA+U-SIC}
%\end{figure}

\subsubsection{Pseudo SIC}
\begin{figure}[hbtp]
 \centering
 \includegraphics[clip,width=1.0\linewidth]{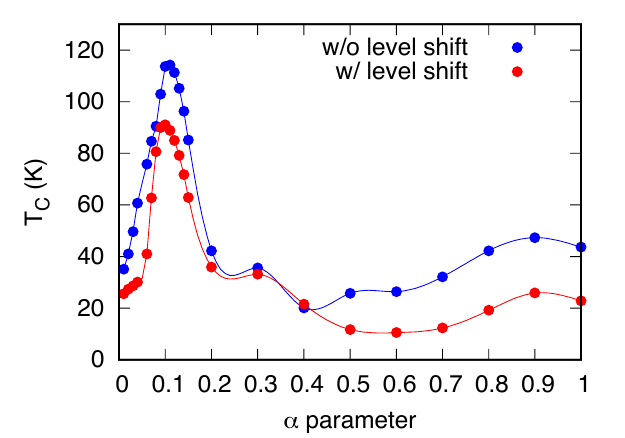}
 \caption{Calculated $T_{\rm C}$ as a function of $\alpha$ using the pSIC scheme}
 \label{fig: Tc pSIC}
\end{figure}

Fig. \ref{fig: Tc pSIC} shows the calculated $T_{\rm C}$ as a function of $\alpha$ parameter using the pSIC scheme with/without artificial level shifts for Ce-$d$ and $f$ states. Without the level shifts, the maximum of the calculated value of $T_{\rm C}^{\rm calc.}\sim 114$ K is obtained for $\alpha=0.14$, which is comparable to $T_{\rm C}^{\rm expt.}$. The same tendency as is the case in the LDA+U-SIC scheme is found that the calculated value of $T_{\rm C}$ is decreased with the level shifts.  The calculation with the level shifts gave the maximum of the calculated value of $T_{\rm C}^{\rm calc.}\sim 91$ K for $\alpha=0.1$. In both cases, we found that the calculated $T_{\rm C}$ comparable to the measured one is attained for $\alpha \sim 0.1$. 

The calculated spin and orbital moments on the Ce site without (with) the level shifts for $\alpha=0.14$ ($\alpha=0.1$) are,  0.84 (0.70) $\mu_{\rm B}$ and -0.65 (-0.51) $\mu_{\rm B}$ for $| -1,1/2\rangle$, 0.89 (0.71) $\mu_{\rm B}$ and -0.07 (-0.07)$\mu_{\rm B}$ for $| 0,1/2\rangle$, respectively, leading to the averaged value of spin and orbital moments as 0.87 (0.71) $\mu_{\rm B}$ and -0.32 (-0.26) $\mu_{\rm B}$.
The calculated value of the total magnetic moment of 0.55 (0.45) $\mu_{\rm B}$ is also comparable to the measured value of $\sim 0.4 \mu_{\rm B}$.

\begin{figure}[hbtp]
	\begin{minipage}{1.0\linewidth}
		\centering
        		\includegraphics[clip,width=1.0\linewidth]{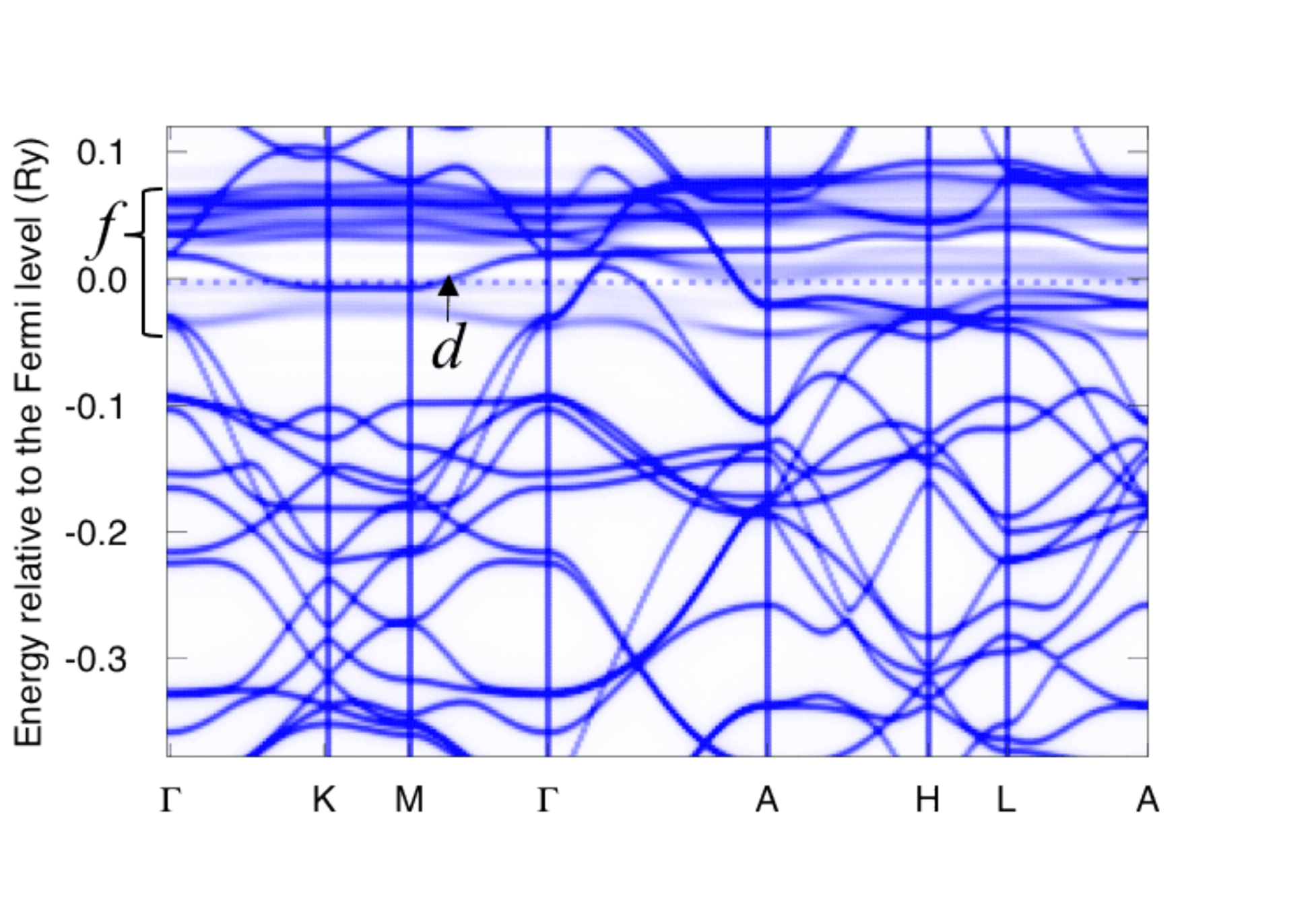}
		(a) without level shifts, $\alpha=0.14$
	\end{minipage}\\
	\begin{minipage}{1.0\linewidth}
		\centering
        		\includegraphics[clip,width=1.0\linewidth]{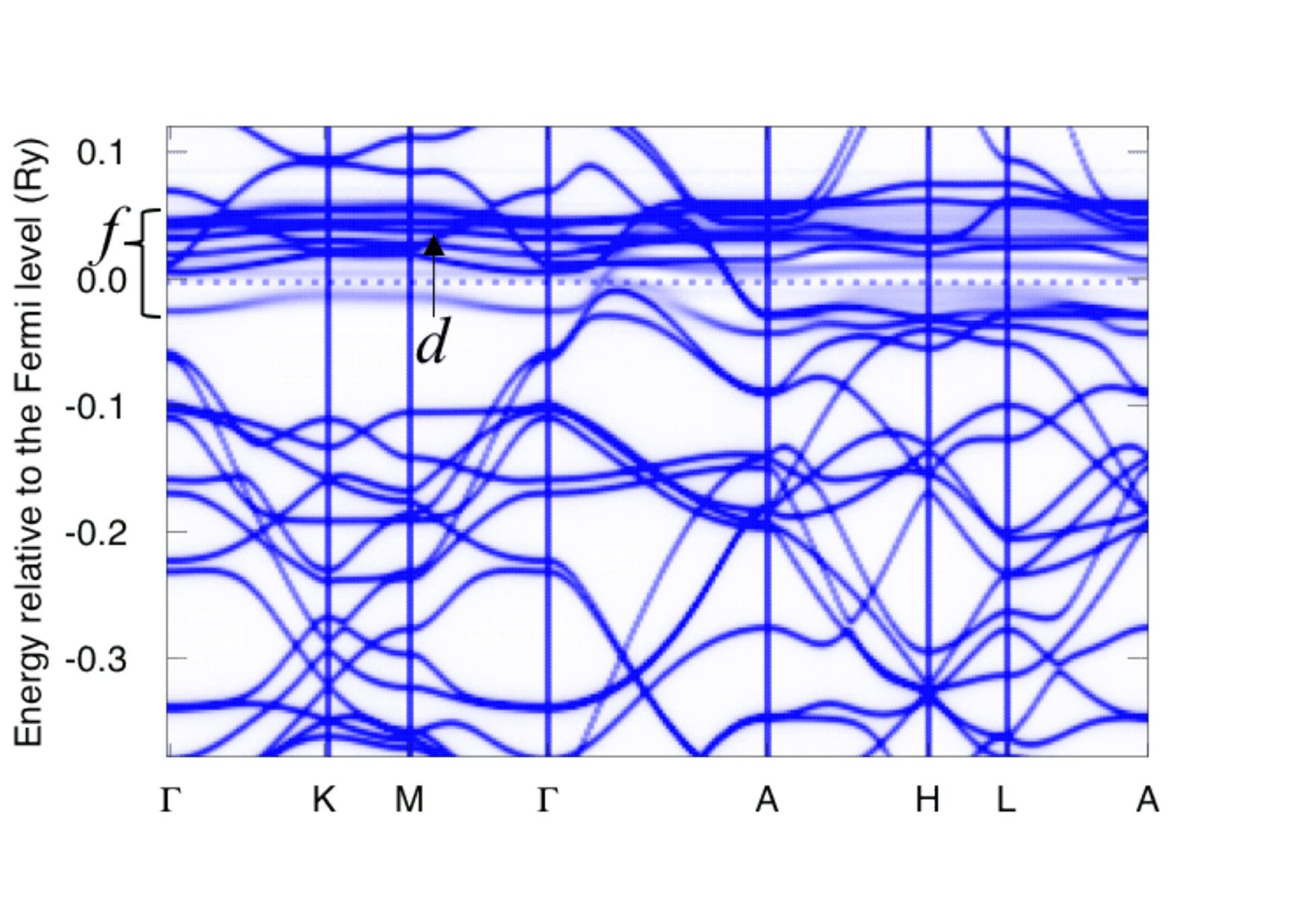}
		(b) with level shifts, $\alpha=0.1$
	\end{minipage}
	\caption{Electronic band structure of Ce$^{3+}$Rh$_{3}$B$_{2}$ for the majority spin channel using the pSIC scheme with $\alpha$ at which the calculated  $T_{\rm C}$ is maximized}
	\label{fig: band pSIC}
	\sublabel{fig: band pSIC without shifts}{(a)}
	\sublabel{fig: band pSIC with shifts}{(b)}	
\end{figure}

Fig. \ref{fig: band pSIC} shows the calculated band dispersion of the majority spin channel for Ce$^{3+}$Rh$_{3}$B$_{2}$ with $\alpha \sim 0.1$, where the calculated $T_{\rm C}$ is maximized, with/without artificial level shifts. The obtained band structure is similar to the one within the LDA+U-SIC scheme as well as the situation where the calculations with the level shifts remedy the topology of the FS as shown in Fig. \ref{fig: band pSIC with shifts}, which presumably is the consequence of the similarity for the correction terms in the both two schemes, Eqs. (\ref{eq: LDA+U-SIC}) and (\ref{eq: pSIC}).

%\begin{figure}[hbtp]
%	\centering
% 	\subfigure[without level shifts, $\alpha=0.14$]{
%       		\centering
%        		\includegraphics[clip,width=1.0\linewidth]{pdf/cerh3b2_ls_psic_3_4_0.14_up_spc.pdf}
%		\label{fig: band pSIC without shifts}
%	}
% 	\subfigure[with level shifts, $\alpha=0.1$]{
%		\centering
%		\includegraphics[clip,width=1.0\linewidth]{pdf/shft_cerh3b2_ls_psic_3_4_0.1_up_spc.pdf}
%		\label{fig: band pSIC with shifts}
%	}
%	\caption{Electronic band structure of Ce$^{3+}$Rh$_{3}$B$_{2}$ for the majority spin channel using the pSIC scheme with $\alpha$ at which the calculated  $T_{\rm C}$ is maximized}
%	\label{fig: band pSIC}
%\end{figure}

\section{Summary and Outlook}
We performed the electronic structure calculations for Ce$^{3+}$Rh$_{3}$B$_{2}$ and estimated the $T_{\rm C}$ via a mapping into an effective spin model using a certain localization correction scheme within a certain constraint for the electronic configuration at the ferromagnetic ground state. The important finding is that the effective interaction between adjacent Ce sites along the $c$-axis is maximized and the corresponding $T_{\rm C}^{\rm calc.}$ is comparable to $T_{\rm C}^{\rm expt.}\sim 120$ K only if the localized $4f$ bands around Ce locate at very shallow level just below the $E_{\rm F}$, which is well consistent with the recent resonant PES experiment. We also confirmed that artificial level shifts for Ce-$d$ and $f$ states are inevitable to reproduce the proper topology of the FS. Within the level shifts combined with the LDA+U-SIC or the pSIC, we obtained the electronic structure that explains the curious properties of CeRh$_{3}$B$_{2}$ such as the small saturation magnetization, the exceptionally high $T_{\rm C}$, and the topology of the FS, simultaneously.

It has been actually known in the community of rare-earth permanent magnets that Ce is basically detrimental to $T_{\rm C}$.
Among the materials family of $4f$-$3d$ intermetallic ferromagnets involving the champion magnet compound Nd$_{2}$Fe$_{14}$B, the $T_{\rm C}$ of Ce$_{2}$Fe$_{14}$B is 424 K \cite{Herbst1991}, which is significantly lower than that of Nd$_{2}$Fe$_{14}$B at 585 K. Alloying Ce in Nd$_{2}$Fe$_{14}$B could potentially enhance the $T_{\rm C}$ by a fabrication of a local environment similar to CeRh$_{3}$B$_{2}$ around the structure made of Nd and B atoms.

\section*{Acknowledgment}
Helpful discussions with H.~Akai, Y.~Kuramoto,
Y.~Onuki, H.~Shishido, Y.~Hirayama are gratefully acknowledged.
MM's work in ISSP is supported by Toyota Motor Corporation.
This work was partly supported by JSPS KAKENHI Grant Number 15K13525
and the Elements Strategy Initiative Center for Magnetic Materials
(ESICMM) under the outsourcing project of the Ministry of Education,
Culture, Sports, Science and Technology (MEXT), Japan.

\bibliographystyle{jpsj}

%\bibliography{ref}

\begin{thebibliography}{10}

\bibitem{Dhar1981}
S.~K. Dhar, S.~K. Malik, and R.~Vijayaraghavan: J. Phys. C {\bfseries 14}
  (1981) L321.

\bibitem{Galatanu2003}
A.~Galatanu, E.~Yamamoto, T.~Okubo, M.~Yamada, A.~Thamizhavel, T.~Takeuchi,
  K.~Sugiyama, Y.~Inada, and Y.~Onuki: J. Phys.: Condens. Matter {\bfseries 15}
  (2003) S2187.

\bibitem{Ku1980}
H.~Ku, G.~Meisner, F.~Acker, and D.~Johnston: Solid State Commun. {\bfseries
  35} (1980) 91 .

\bibitem{Malik1982}
S.~K. Malik, S.~K. Dhar, R.~Vijayaraghavan, and W.~E. Wallace: J. Appl. Phys.
  {\bfseries 53} (1982) 8074.

\bibitem{Malik1983}
S.~Malik, R.~Vijayaraghavan, W.~Wallace, and S.~Dhar: J. Magn. Magn. Mater.
  {\bfseries 37} (1983) 303 .

\bibitem{Sampathkumaran1985}
E.~V. Sampathkumaran, G.~Kaindl, C.~Laubschat, W.~Krone, and G.~Wortmann: Phys.
  Rev. B {\bfseries 31} (1985) 3185.

\bibitem{Yaouanc1998}
A.~Yaouanc, P.~Dalmas~de R\'eotier, J.-P. Sanchez, T.~Tschentscher, and
  P.~Lejay: Phys. Rev. B {\bfseries 57} (1998) R681.

\bibitem{Alonso1998}
J.~Alonso, J.~Boucherle, F.~Givord, J.~Schweizer, B.~Gillon, and P.~Lejay: J.
  Magn. Magn. Mater. {\bfseries 177-181} (1998) 1048 .

\bibitem{Ebihara2000}
T.~Ebihara, Y.~Inada, M.~Murakawa, S.~Uji, C.~Terakura, T.~Terashima,
  E.~Yamamoto, Y.~Haga, Y.~^^c5^^8cnuki, and H.~Harima: J. Phys. Soc. Jpn
  {\bfseries 69} (2000) 895.

\bibitem{Okubo2003}
T.~Okubo, M.~Yamada, A.~Thamizhavel, S.~Kirita, Y.~Inada, R.~Settai, H.~Harima,
  K.~Takegahara, A.~Galatanu, E.~Yamamoto, and Y.~^^c5^^8cnuki: J. Phys.:
  Condens. Matter {\bfseries 15} (2003) L721.

\bibitem{Givord2004}
F.~Givord, J.-X. Boucherle, E.~Leli{\`{e}}vre-Berna, and P.~Lejay: J. Phys.:
  Condens. Matter {\bfseries 16} (2004) 1211.

\bibitem{Kishimoto2004}
Y.~Kishimoto, Y.~Kawasaki, and T.~Ohno: J. Phys. Soc. Jpn {\bfseries 73} (2004)
  1970.

\bibitem{Yamada2004}
M.~Yamada, Y.~Obiraki, T.~Okubo, T.~Shiromoto, Y.~Kida, M.~Shiimoto, H.~Kohara,
  T.~Yamamoto, D.~Honda, A.~Galatanu, Y.~Haga, T.~Takeuchi, K.~Sugiyama,
  R.~Settai, K.~Kindo, S.~K.~Dhar, H.~Harima, and Y.~^^c5^^8cnuki: J. Phys.
  Soc. Jpn {\bfseries 73} (2004) 2266.

\bibitem{Imada2007}
S.~Imada, A.~Yamasaki, M.~Tsunekawa, A.~Higashiya, A.~Sekiyama, H.~Sugawara,
  H.~Sato, and S.~Suga: J. Electron. Spectrosc. Relat. Phenom. {\bfseries
  156-158} (2007) 436 .

\bibitem{Givord2007aug}
F.~Givord, J.-X. Boucherle, R.-M. Gal{\'{e}}ra, G.~Fillion, and P.~Lejay: J.
  Phys.: Condens. Matter {\bfseries 19} (2007) 356208.

\bibitem{Givord2007nov}
F.~Givord, J.-X. Boucherle, A.~P. Murani, R.~Bewley, R.-M. Gal{\'{e}}ra, and
  P.~Lejay: J. Phys.: Condens. Matter {\bfseries 19} (2007) 506210.

\bibitem{Raymond2010}
S.~Raymond, J.~Panarin, F.~Givord, A.~P. Murani, J.~X. Boucherle, and P.~Lejay:
  prb {\bfseries 82} (2010) 094416.

\bibitem{Ito2014}
M.~Ito, K.~Suzuki, T.~Tadenuma, R.~Nagayasu, Y.~Sakurai, Y.~Onuki,
  E.~Nishibori, and M.~Sakata: J. Phys.: Conf. Ser. {\bfseries 502} (2014)
  012018.

\bibitem{Takegahara1985}
K.~Takegahara, H.~Harima, and T.~Kasuya: J. Phys. Soc. Jpn {\bfseries 54}
  (1985) 4743.

\bibitem{Eriksson1989a}
O.~Eriksson, B.~Johansson, M.~S.~S. Brooks, H.~L. Skriver, and J.~Sj\"ostr\"om:
  Phys. Rev. B {\bfseries 40} (1989) 5270.

\bibitem{Harima2004}
H.~Harima and K.~Takegahara: J. Magn. Magn. Mater. {\bfseries 272-276} (2004)
  475 .

\bibitem{Kono2006}
H.~N.~Kono and Y.~Kuramoto: J. Phys. Soc. Jpn {\bfseries 75} (2006) 084706.

\bibitem{Yamauchi2010}
K.~Yamauchi, A.~Yanase, and H.~Harima: J. Phys. Soc. Jpn {\bfseries 79} (2010)
  044717.

\bibitem{Moruzzi1978}
V.~L. Moruzzi, J.~F. Janak, and A.~R. Williams: {\em Calculated electronic
  properties of metals} (Pergamon Press Inc., 1978).

\bibitem{Anisimov1997}
V.~I. Anisimov, F.~Aryasetiawan, and A.~I. Lichtenstein: J. Phys.: Condens.
  Matter {\bfseries 9} (1997) 767.

\bibitem{Perdew1981}
J.~P. Perdew and A.~Zunger: Phys. Rev. B {\bfseries 23} (1981) 5048.

\bibitem{Czyzyk1994}
M.~T. Czy\ifmmode~\dot{z}\else \.{z}\fi{}yk and G.~A. Sawatzky: Phys. Rev. B
  {\bfseries 49} (1994) 14211.

\bibitem{Dudarev1998}
S.~L. Dudarev, G.~A. Botton, S.~Y. Savrasov, C.~J. Humphreys, and A.~P. Sutton:
  Phys. Rev. B {\bfseries 57} (1998) 1505.

\bibitem{Shick1999}
A.~B. Shick, A.~I. Liechtenstein, and W.~E. Pickett: Phys. Rev. B {\bfseries
  60} (1999) 10763.

\bibitem{Seo2007}
D.-K. Seo: Phys. Rev. B {\bfseries 76} (2007) 033102.

\bibitem{Filippetti2003}
A.~Filippetti and N.~A. Spaldin: Phys. Rev. B {\bfseries 67} (2003) 125109.

\bibitem{Pemmaraju2007}
C.~D. Pemmaraju, T.~Archer, D.~S\'anchez-Portal, and S.~Sanvito: Phys. Rev. B
  {\bfseries 75} (2007) 045101.

\bibitem{Filippetti2009}
A.~Filippetti and V.~Fiorentini: Eur. Phys. J. B {\bfseries 71} (2009) 139.

\bibitem{Filippetti2011}
A.~Filippetti, C.~D. Pemmaraju, S.~Sanvito, P.~Delugas, D.~Puggioni, and
  V.~Fiorentini: Phys. Rev. B {\bfseries 84} (2011) 195127.

\bibitem{Liechtenstein1987}
A.~Liechtenstein, M.~Katsnelson, V.~Antropov, and V.~Gubanov: J. Magn. Magn.
  Mater. {\bfseries 67} (1987) 65 .

\bibitem{Cococcioni2005}
M.~Cococcioni and S.~de~Gironcoli: Phys. Rev. B {\bfseries 71} (2005) 035105.

\bibitem{Faulkner1980}
J.~S. Faulkner and G.~M. Stocks: Phys. Rev. B {\bfseries 21} (1980) 3222.

\bibitem{Herbst1991}
J.~F. Herbst: Rev. Mod. Phys. {\bfseries 63} (1991) 819.

\end{thebibliography}

\end{document}